\documentclass[sigconf]{acmart}

\AtBeginDocument{%
  }

\copyrightyear{2025} 
\acmYear{2025} 
\setcopyright{acmlicensed}
\acmConference[CHI '25]{CHI Conference on Human Factors in Computing Systems}{April 26-May 1, 2025}{Yokohama, Japan}
\acmBooktitle{CHI Conference on Human Factors in Computing Systems (CHI '25), April 26-May 1, 2025, Yokohama, Japan}
\acmDOI{10.1145/3706598.3713739}
\acmISBN{979-8-4007-1394-1/25/04}

\newcommand{\mytextcolor}[2]{#2}

\begin{document}

\title{GazeSwipe: Enhancing Mobile Touchscreen Reachability through Seamless Gaze and Finger-Swipe Integration}

\author{Zhuojiang Cai}
\authornote{Both authors contributed equally to this research.}
\orcid{0009-0005-3404-118X}
\affiliation{%
  \institution{State Key Lab. of VR Technology and Systems\\Beihang University}
  \city{Beijing}
  \country{China}
}
\email{caizhuojiang@buaa.edu.cn}

\author{Jingkai Hong}
\authornotemark[1]
\orcid{0009-0001-0568-3360}
\affiliation{%
  \institution{State Key Lab. of VR Technology and Systems\\Beihang University}
  \city{Beijing}
  \country{China}
}
\email{jingkai_hong@buaa.edu.cn}

\author{Zhimin Wang}
\orcid{0000-0001-5089-977X}
\affiliation{%
  \institution{State Key Lab. of VR Technology and Systems\\Beihang University}
  \city{Beijing}
  \country{China}
}
\email{zm.wang@buaa.edu.cn}

\author{Feng Lu}
\authornote{Feng Lu is the corresponding author.}
\orcid{0000-0001-9064-7964}
\affiliation{%
  \institution{State Key Lab. of VR Technology and Systems\\Beihang University}
  \city{Beijing}
  \country{China}
}
\email{lufeng@buaa.edu.cn}

\renewcommand{\shortauthors}{Cai et al.}

\begin{abstract}

    Smartphones with large screens provide users with increased display and interaction space but pose challenges in reaching certain areas with the thumb when using the device with one hand. To address this, we introduce GazeSwipe, a multimodal interaction technique that combines eye gaze with finger-swipe gestures, enabling intuitive and low-friction reach on mobile touchscreens.
    Specifically, we design a gaze estimation method that eliminates the need for explicit gaze calibration. Our approach also avoids the use of additional eye-tracking hardware by leveraging the smartphone's built-in front-facing camera. Considering the potential decrease in gaze accuracy without dedicated eye trackers, we use finger-swipe gestures to compensate for any inaccuracies in gaze estimation. Additionally, we introduce a user-unaware auto-calibration method that improves gaze accuracy during interaction.
    Through extensive experiments on smartphones and tablets, we compare our technique with various methods for touchscreen reachability and evaluate the performance of our auto-calibration strategy. The results demonstrate that our method achieves high success rates and is preferred by users. The findings also validate the effectiveness of the auto-calibration strategy.

\end{abstract}


\begin{teaserfigure}
  \includegraphics[width=\textwidth]{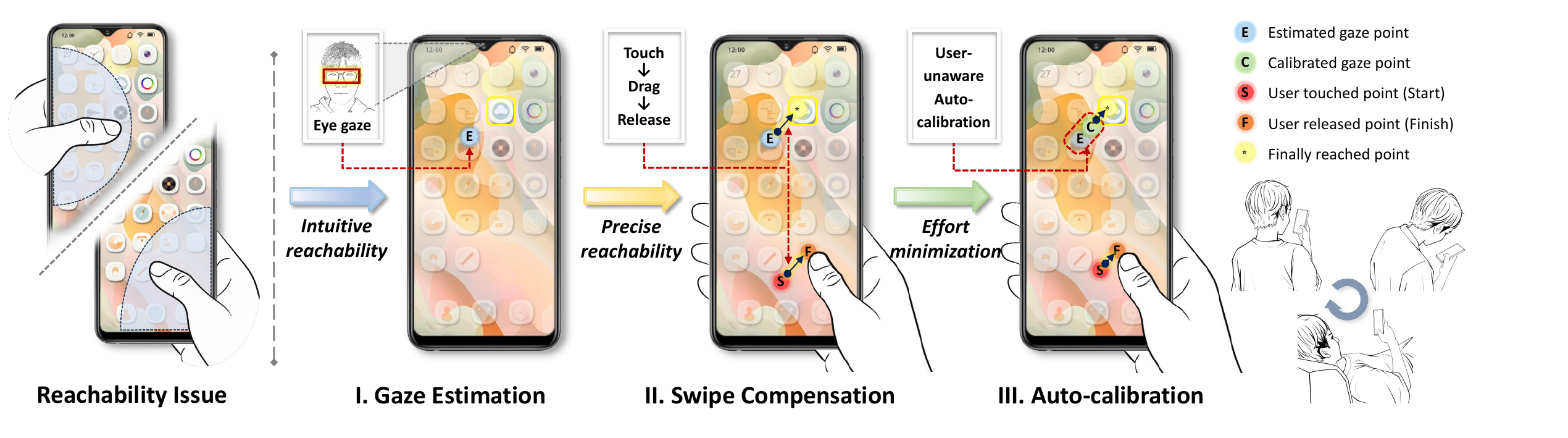}
  \caption{To address the issue of thumb reachability on smartphones and tablets when held with one hand, we introduce the gaze-based pointing, enabling users to intuitively reach any position by simply gazing at the target. By combining this with the finger-touch input, users can precisely interact with targets through swipe gestures. Additionally, we
  introduce a user-unaware auto-calibration method that 
  eliminates the need for explicit gaze calibration, enhancing gaze accuracy during use and making interactions more seamless and efficient.}
  \Description{The basic concept of the proposed method, starting from the issue of one-handed reachability on smartphones. Initially, gaze estimation is introduced to achieve full-screen reachability, followed by precise target selection through touch compensation. Finally, the accuracy of gaze guidance is enhanced by incorporating our auto-calibration strategy.}
  \label{fig:teaser}
\end{teaserfigure}


\maketitle

\section{Introduction}

In recent years, the rapid advancement of smartphone and tablet technology has been characterized by a significant increase in screen sizes. While these larger displays have undoubtedly enriched user experiences by providing more space for content presentation and interaction, they have also introduced a series of ergonomic challenges, particularly concerning one-handed usage \cite{eardley_understanding_2017, kim_interaction_2012, ikematsu_investigating_2020}. As screen sizes approach or exceed the 5.5-inch threshold, users find it increasingly difficult to reach certain areas of the screen with one hand \cite{voelker_headreach_2020}. This often requires frequent adjustments to the grip, making operations more cumbersome and time-consuming, and increasing the risk of accidental drops.

Imagine you are holding your smartphone with one hand while the other is occupied by carrying a bag, holding onto a bus or subway handrail, or steering a shopping cart, and you need to tap a button at the top of the screen. Alternatively, picture yourself walking through the office, one hand holding a coffee cup while the other operates a tablet, or even writing on a whiteboard with one hand while trying to use a tablet with the other. In these situations, reaching distant parts of the screen becomes difficult and unstable. These scenarios highlight the reachability issues of large touchscreens when used with one hand, emphasizing the need for a more convenient and efficient solution.

To address the reachability issue, several solutions have been proposed. Screen Transform Techniques, such as Sliding Screen \cite{kim_interaction_2012}, TiltReduction \cite{chang_understanding_2015}, and One-handed Mode features on various smartphones \cite{huawei_one-handed_2019}, adjust the user interface by scaling or shifting it to the lower areas of the screen, bringing content within the thumb’s comfortable reach. While these methods can enhance reachability and improve grip stability, they come at the cost of reducing the usable screen area. Scaling the interface often results in smaller elements, leading to imprecise operations, while shifting the interface to display only a part of the content makes it inconvenient for continuous operations across different regions of the screen. Moreover, since the thumb's reachable area remains fixed regardless of screen size, the required degree of scaling or shifting increases with larger displays, making these techniques impractical for tablets.

Another category of solutions, Cursor-based Techniques, such as BezelCursor \cite{li_bezelcursor_2013}, ForceRay \cite{corsten_forceray_2019}, and HeadReach \cite{voelker_headreach_2020}, enable users to control a virtual cursor to reach targets anywhere on the screen via touch. While these strategies avoid the limitations of Screen Transform Techniques, their interaction experiences are often less intuitive and require long thumb movements, making the interaction more effortful and time-consuming.

Given that gaze information naturally and intuitively reflects user intent, increasing research is integrating it into mobile device interactions \cite{nagamatsu_mobigaze_2010, dybdal_gaze_2012, khamis_past_2018}. For example, gaze has been used to enhance mobile voice assistants, improve text editing efficiency, and facilitate cross-device collaboration \cite{mayer_enhancing_2020, sindhwani_retype_2019, voelker_gazeconduits_2020}. CursorShift \cite{pfeuffer_gaze_2016} achieved gaze and touch interaction on tablets in a controlled lab environment. However, this method requires expensive external eye-tracking devices and additional explicit calibration processes, making it impractical for everyday use on mobile devices and insufficient for effectively solving reachability problems. 
Recently, Apple has also introduced gaze tracking on iPhones and iPads as an accessibility feature for people with disabilities, allowing users to gaze at a target and dwell to confirm interactions \cite{butler_apple_2024}. However, this feature requires calibration before use, and after calibration, changes in head position may lead to inaccurate performance. Moreover, the reliance on gaze alone for unimodal interaction makes it susceptible to the Midas touch problem. Therefore, it cannot serve as a viable solution for the everyday reachability issue.

To this end, we propose GazeSwipe, a multimodal interaction technique that combines eye gaze with finger-swipe gestures to enhance reachability on smartphones and tablets in an intuitive and low-effort manner.
With GazeSwipe, users simply gaze at the target and use a thumb swipe to adjust the gaze cursor and complete the interaction. Our approach offers three key advantages: 1) it requires no additional hardware, utilizing the front-facing camera for gaze estimation; 2) it employs a user-unaware auto-calibration strategy, eliminating the need for explicit gaze calibration and improving gaze accuracy during use; and 3) 
Since gaze is fast and intuitive, while finger-swipe is subtle and effortless, the combination delivers a seamless and low-friction interaction experience for everyday use on smartphones and tablets.

We conduct two user studies to evaluate our interaction technique. In the first study, we examine the impact of different calibration strategies on GazeSwipe. The results show that our auto-calibration strategy significantly improves gaze estimation accuracy while eliminating the annoying explicit calibration process and enhancing robustness to head movements in everyday use. In the second study, we compared GazeSwipe with other interaction techniques addressing reachability issues on smartphones and tablets. The results demonstrate that our method achieves high success rates and is prefered by users on both devices. Specifically on tablets, our method outperforms others across all subjective and objective metrics, except for a moderately longer completion time.

In summary, this paper makes the following contributions:

\begin{itemize}
    \item We propose GazeSwipe, a novel multimodal interaction technique that integrates eye gaze with finger swipes, providing an intuitive and low-friction method for enhancing reachability on smartphones and tablets.
    \item We propose a user-unaware auto-calibration method that eliminates the need for explicit gaze calibration while enhancing gaze accuracy during use, thereby enabling smooth interactions.
    \item We conducted extensive user studies to evaluate the auto-calibration strategy and compare GazeSwipe with other reachability techniques, demonstrating its promising performance on both smartphones and tablets.
\end{itemize}

\section{Related Work}

\subsection{Reachability Techniques on Mobile Devices}

As smartphones continue to increase in screen size, there is a growing challenge of thumb reachability to certain areas of the screen while holding the device with one hand. To address this issue and enable users to interact with targets in these unreachable areas more conveniently, researchers have explored various reachability techniques. We categorize these techniques into screen transform techniques and cursor techniques for discussion.

\subsubsection{Screen Transform Techniques}

The screen transform technique involves resizing or moving screen regions to bring unreachable areas closer to the thumb. Many smartphones already implement such features, such as One-handed Mode on Android devices \cite{huawei_one-handed_2019}. This mode is activated by swiping up from one of the bottom corners, causing the interface to shrink towards that corner, thereby allowing the thumb to reach previously unreachable areas. Similarly, MovingScreen \cite{tsai_movingscreen_2016} and Sliding Screen \cite{kim_interaction_2012} utilize edge swipes to to move the screen closer to the thumb. ThumbSpace \cite{baranauskas_thumbspace_2007} creates a pop-up view at the thumb touch position to display the entire screen, while TapTap \cite{roudaut_taptap_2008} shows a portion of the screen near the thumb touch position in the pop-up view. Chang et al. \cite{chang_understanding_2015} proposed tilt-based solutions for reachability issues, including TiltSlide, which shifts the screen towards the tilt direction, and TiltReduction, which reduces the screen size when the device is tilted. Le et al. \cite{le_investigating_2016} suggested moving the interface by sliding the index finger on a touchpad located on the back of the device.

However, these methods often shrink or partially display the interface, making it harder to accurately tap targets and affecting display visibility. Tilt-controlled approaches may be impractical in certain postures, such as when lying down. Additionally, methods using back touchpads require additional hardware.

\subsubsection{Cursor Techniques}

Another approach, besides moving the screen closer to the thumb, is to control the cursor to interact with unreachable screen areas. BezelCursor \cite{li_bezelcursor_2013} and ExtendedThumb \cite{lai_extendedthumb_2015} trigger cursor mode through bezel swipes or double-taps, enabling users to drag the cursor to the desired target. TiltCursor \cite{chang_understanding_2015} triggers cursor mode when the device is tilted. Ikeda et al. \cite{ikeda_hover-based_2021} proposed a hover-based method for cursor manipulation. CornerSpace \cite{yu_rapid_2013} permits users to position a remote cursor in unreachable screen zones for swift access. ForceRay \cite{corsten_forceray_2019} utilizes the force-sensitive screen to adjust cursor distance from the thumb based on pressure intensity. Dual-Surface Input \cite{yang_dual-surface_2009} employs an extra touchpad on the device's back to select targets beyond the thumb's reach. HeadReach \cite{voelker_headreach_2020} enables cursor control via head movements. Esteves et al. \cite{esteves_one-handed_2022} allow users to select predefined targets outside the screen by specifying sliding trajectories on the screen.

Although these cursor-based techniques can help users accurately reach targets, the long thumb movement distances may lead to user fatigue. Additionally, the use of force input, head posture, and custom sliding trajectories to assist in addressing reachability issues may not be entirely intuitive, requiring a learning curve.

\subsection{Gaze Estimation Techniques}

Gaze estimation technique involves predicting the point of gaze (PoG) or gaze direction from images or videos captured by a camera. These methods can be categorized into model-based and appearance-based techniques.

\subsubsection{Model-based Techniques}

Model-based techniques estimate gaze by fitting a 3D eye model from images. Some methods utilize specialized RGB-D cameras \cite{xiong_eye_2014,sun_real_2015,funes_mora_geometric_2014} or infrared cameras \cite{fitzgibbon_point_2012} to obtain 3D facial landmarks, 3D pupil location, iris contour, and corneal reflection, enabling more accurate fitting of the eye model. However, these methods are not suitable for use on mobile devices due to the requirement for additional equipment.

Other methods utilize regular RGB cameras \cite{valenti_combining_2012,lu_estimating_2016,wang_real_2017,ververas_3dgazenet_2023} to capture images for fitting the eye model. They use detected 2D facial landmarks to fit a 3D face model to obtain head pose and detect the 2D pupil center \cite{valenti_combining_2012} or iris contour \cite{lu_estimating_2016}. Wang et al. \cite{wang_real_2017} proposed a deformable eye-face model, while Ververas et al. \cite{ververas_3dgazenet_2023} proposed using neural network regression to predict dense 3D eye meshes, enhancing estimation robustness. Nevertheless, these methods are generally less robust in unconstrained environments.

\subsubsection{Appearance-based Techniques}
Appearance-based techniques directly map appearance images to 2D gaze points or 3D gaze directions, making them suitable for use on mobile devices. For estimating 2D gaze points, Krafka et al. \cite{krafka_eye_2016} introduced a mobile device-compatible gaze estimation technique, utilizing the GazeCapture dataset to train a convolutional neural network. This network takes cropped eye and face images as input, eliminating the need for additional head pose information. Bao et al. \cite{bao_adaptive_2021} enhanced gaze estimation accuracy by adaptively integrating binocular features and guiding eye feature extraction with facial cues. Huynh et al. \cite{huynh_imon_2021} proposed iMon, a mobile-specific gaze tracking system, which significantly improves estimation precision by addressing gaze uncertainty due to minor eye movements, refining eye images to reduce motion blur, and correcting individual focus differences. Park et al. \cite{park_towards_2020} proposed a method integrating visual saliency to address accuracy challenges in traditional eye-tracking caused by individual differences and data scarcity.

For estimating 3D gaze directions, several methods have been proposed. Fischer et al. \cite{ferrari_rt-gene_2018} introduced RT-GENE, a real-time 3D gaze estimation system designed for natural environments, leveraging advanced image processing and feature extraction techniques. Kellnhofer et al. \cite{kellnhofer_gaze360_2019} presented Gaze360, which combines a 360° panoramic camera setup with multiple SLR cameras to collect comprehensive data for accurate 3D gaze direction estimation. Zhang et al. \cite{zhang_eth-xgaze_2020} developed ETH-XGaze, a large-scale dataset and accompanying techniques tailored for precise 3D gaze estimation under extreme head pose and gaze variations. Cheng et al. \cite{cheng_gaze_2022} proposed GazeTR, one of the first methods to adopt the vision transformer architecture for gaze estimation, enhancing accuracy through a hybrid approach that integrates CNN feature extraction. Nagpure et al. \cite{nagpure_searching_2023} introduced GazeNAS, employing neural architecture search to create lightweight and efficient 3D gaze estimation models.

Despite significant progress, appearance-based techniques still cannot provide precise gaze estimation for fine-grained interaction on mobile devices in unconstrained environments. In our work, we employ a lightweight 2D gaze estimation method that enables direct usage on mobile devices without requiring gaze point projection.

\subsection{Gaze-based Interaction Techniques}

In this paper, we propose a gaze-based dual-modal interaction technique that combines gaze and touch inputs on mobile touchscreens. This section reviews gaze-only and gaze-based multi-modal interaction methods, with a particular focus on those on mobile devices.

\subsubsection{Gaze-only Interaction}

Gaze-only interaction on computer displays and mobile touchscreens has been explored in many studies. Previous work has classified gaze-based interactions into two categories: implicit and explicit \cite{namnakani_comparing_2023, lei_end--end_2024}. Implicit interactions trigger actions based on the user's passive gaze behavior, such as triggering page scrolling by analyzing gaze trajectories during reading \cite{lei_dynamicread_2023}. Explicit interactions, more relevant to our work, rely on intentional gaze movements to trigger actions. Common techniques in explicit gaze interaction include dwell time, pursuit, and gaze gestures \cite{drewes_eye-gaze_2007, vidal_pursuits_2013, namnakani_comparing_2023}. For instance, Apple’s eye-tracking accessibility feature on iOS uses dwell time to trigger interactions \cite{butler_apple_2024}.

Although gaze-only systems hold significant potential, they face critical challenges. One major issue is the Midas touch problem, where unintended gaze fixations inadvertently trigger actions on the screen, leading to false inputs. In addition, some methods may cause user fatigue, making them less suitable for prolonged use.

\subsubsection{Gaze + Voice / Motion / Mouse}

Other studies have explored combining gaze with other modalities for interaction on mobile touchscreens \cite{lei_end--end_2024}. Mayer et al. \cite{mayer_enhancing_2020} introduced WorldGaze, a pioneering technology aimed at enhancing mobile voice assistants through gaze interaction. Zhao et al. \cite{zhao_eyesaycorrect_2022} proposed a hands-free text correction method that combines gaze and voice on smartphones. Kong et al. \cite{kong_eyemu_2021} utilized both gaze and hand motion for controlling smartphone interfaces. These studies collectively underscore the crucial role of user visual attention information provided by gaze in mobile device interactions.

Some gaze and motion-based interaction techniques in XR employ auto-calibration methods. Kytö et al. \cite{kyto_pinpointing_2018} used gaze in AR headsets, refining interactions through head or hand movements and collecting temporal interaction samples for calibration. Similarly, Sidenmark et al. \cite{sidenmark_gaze_2019} explored gaze-hand coordination calibration strategies in VR headsets. While these strategies have proven successful in XR, they have not been applied to mobile touchscreen scenarios, where optimizations such as considering changes in head posture are important. 

Additionally, Sugano et al. \cite{forsyth_incremental_2008} explored gaze estimation with auto-calibration in desktop PC environments, considering head movement. While there are some similarities with our approach, key differences arise from the different devices used (smartphones and tablets vs. desktop PCs). Additionally, our auto-calibration strategy collects data from finger swipe release positions (vs. from mouse clicks \cite{forsyth_incremental_2008}), which is specifically tailored to our interaction method. This combination of gaze, finger swipe, and auto-calibration ensures smooth and precise interactions for everyday use.

\subsubsection{Gaze + Touch}

The integration of gaze and touch has been increasingly explored in recent research. Rivu et al. \cite{rivu_gazentouch_2020} proposed a method for text editing that combines gaze and touch. Pfeuffer et al. \cite{pfeuffer_multi-user_2021} explored gaze and touch interaction for multi-user touchscreen collaboration. Voelker et al. \cite{voelker_gazeconduits_2020} utilized gaze and touch for cross-device collaboration. Rivu et al. \cite{rivu_gazebutton_2019} introduced an enhanced button interaction method based on gaze. Some works \cite{khamis_gtmopass_2017,khamis_user-centred_2022} employ the combination of gaze and touch for authentication on mobile devices.

While these studies cover a variety of applications, some research has specifically explored the use of gaze with touch swipes for refinement across different devices. Stellmach et al. \cite{stellmach_look_2012} proposed a method for controlling a distant display using gaze and a handheld touchscreen. Pfeuffer et al. \cite{pfeuffer_gaze-shifting_2015} demonstrated interactions on tablets using a combination of pen, touch, and gaze. The work most closely related to ours is CursorShift \cite{pfeuffer_gaze_2016}, which uses a Tobii EyeX tracker for gaze estimation and combines it with touch input on tablet devices. However, these approaches focus on different contexts from ours. While they demonstrate how gaze and touch can collaborate for interaction, the need for a large and fixed eye tracker, along with additional calibration processes, makes these methods impractical for smartphones and tablets in everyday mobile scenarios and for addressing reachability challenges.

\section{Motivation and Challenges}

Given the limitations of Screen Transform Techniques---such as wasted screen space, imprecise operations, and difficulty in extending to tablets---a more efficient and low-friction Cursor-based Technique might be a better reachability solution on both smartphones and tablets. Based on this motivation, we aim to develop a gaze cursor technique that intuitively and efficiently overcomes the reachability issue. However, this technique still faces the following three challenges:

\textit{Q1: Is it possible to implement gaze cursor interaction without additional hardware or a calibration process?} Previous work has attempted to implement gaze and touch interaction on tablets, but it relied on external, expensive eye trackers and required a cumbersome 9-point calibration process \cite{pfeuffer_gaze_2016}. This limits its practicality for everyday mobile use and its ability to effectively address the reachability issue. Can we instead leverage the front-facing camera of a mobile device, without requiring calibration, to provide an effortless and effective reachability solution?

\textit{Q2: If there is no calibration process, how can the gaze cursor be made more accurate and stable?} The gaze estimation results are visualized as a gaze cursor, and its accuracy directly affects the distance the user needs to adjust the cursor. Without additional calibration processes and relying solely on the front-facing camera, the accuracy of gaze estimation is typically low. So, how can we improve the accuracy and stability of the gaze estimation results?

\textit{Q3: How can we design a gaze cursor-based interaction method that is natural and seamless for users?} Even with professional external eye trackers, gaze cursors often lack the precision required for fine-grained interactions. Thus, despite improvements in gaze estimation accuracy and stability using the front-facing camera, users may still need to adjust the cursor. Given this challenge, how can we develop a smooth and user-friendly interaction method?

\begin{figure*}[tb]
 \centering 
 \includegraphics[width=\linewidth]{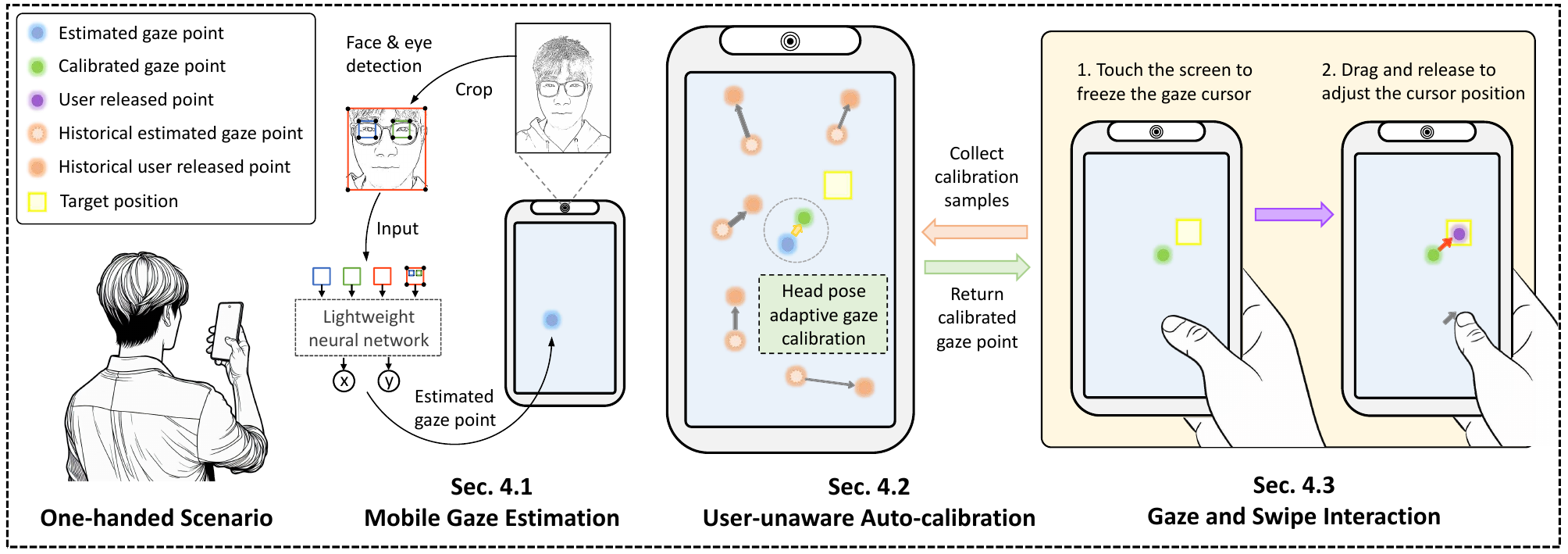}
 \caption{Overview of the GazeSwipe method. Our approach is designed to tackle the reachability challenges encountered in one-handed mobile touchscreen usage scenarios. It involves feeding facial images captured by the front-facing camera into a lightweight neural network to estimate the gaze point and display a cursor accordingly. Additionally, the system performs user-unaware auto-calibration based on historical interaction samples, improving gaze accuracy over time. To complete interactions, users gaze at the target and fine-tune the gaze cursor position using a ``touch-drag-release'' approach.}
 \Description{Overview of the method proposed in this paper, which is divided into three main parts corresponding to Sections 4.1 to 4.3: mobile gaze estimation, user-unaware auto-calibration, and gaze and swipe interaction. Our approach is designed to tackle the reachability challenges encountered in one-handed mobile touchscreen usage scenarios. It involves feeding facial images captured by the front-facing camera into a lightweight neural network to estimate the gaze point and display a cursor accordingly. Additionally, the system performs user-unaware auto-calibration based on historical interaction samples, improving gaze accuracy over time. To complete interactions, users gaze at the target and fine-tune the gaze cursor position using a ``touch-drag-release'' approach.}
 \label{fig:method}
\end{figure*}

\section{GazeSwipe}

Building on the motivation and challenges outlined above, we propose GazeSwipe, an intuitive and low-effort gaze-and-swipe interaction technique that requires no calibration and can be used in everyday scenarios. Our approach is presented in three parts: mobile gaze estimation, user-unaware auto-calibration, and gaze-and-swipe interaction.

\subsection{Mobile Gaze Estimation}

GazeSwipe relies on the gaze estimation method to estimate the user's real-time gaze point on the screen, enabling quick and accurate pointing of the intended interaction target. To achieve a plug-and-play functionality without requiring calibration, we employ a deep learning approach that estimates the gaze point directly from facial images captured by the device's front-facing camera.

Specifically, our approach employs a lightweight neural network \cite{bao_adaptive_2021} to estimate the gaze point coordinates on the screen plane based on RGB images, with the camera as the origin and distances measured in centimeters. This neural network operates in real-time on most mainstream smartphones and achieves competitive performance on the GazeCapture \cite{krafka_eye_2016} and MPIIFaceGaze \cite{zhang_mpiigaze_2017} datasets, a gaze error of 3.9 cm on the MPIIFaceGaze dataset. On our experimental devices, the accuracy is 2.9 cm for the phone and 3.5 cm for the tablet, with an average frame rate of 10-15 fps.

The gaze point is then transformed into screen coordinates according to the camera and screen parameters, and a one-euro filter is applied to eliminate gaze jitter. Let \( \mathcal{I} \) represent the image captured by the front camera, and \( G_{E} \) represent the estimated gaze point in the screen coordinate system. Our gaze estimation process can be formulated as:
\begin{equation}
G_{E} = H(T \cdot f_{\theta}(\mathcal{I}))
\end{equation}
where \( f_{\theta}(\cdot) \) is the neural network, and \( \theta \) represents its parameters. The transformation matrix \( T \) converts the coordinates from the camera-origin system to the screen-origin system, and \( H(\cdot) \) represents the filter used to smooth the estimated gaze points.

However, the estimated gaze point alone is insufficient for precise pointing of fine-grained interaction elements. Therefore, we introduce an auto-calibration method to further enhance gaze accuracy, which is detailed in Section \ref{sec:calibration}, and an interaction method that combines gaze with swipe gestures, described in Section \ref{sec:interaction}.

\subsection{User-Unaware Auto-Calibration}
\label{sec:calibration}
 
Existing gaze tracking techniques typically require individual calibration before use. In conventional gaze interaction techniques, an explicit 9-point calibration strategy is one of the mainstream methods. Users are required to sequentially gaze at 9 target points displayed on the screen for a few seconds each before using the device. This process captures eye parameters or adjusts the mapping relationship to improve the accuracy of the gaze estimation. However, this explicit calibration process is cumbersome and significantly reduces the user experience. Furthermore, in mobile touchscreen use cases, the user may frequently move, causing them to gradually deviate from the initial calibration environment during subsequent use. These deviations may be due to changes in head posture, variations in the relative position between the face and the camera, or changes in lighting conditions. As a result, calibration information may become invalid, leading to increased gaze estimation errors over time. While recalibration can mitigate these deviations, frequent calibration is impractical in daily use.

To overcome the limitations of explicit gaze calibration strategies on mobile touchscreens, we propose a user-unaware auto-calibration method. This method is based on the insight that the interaction process itself can serve as the calibration process, enabling implicit, dynamic, and incremental gaze calibration during interactions. Our auto-calibration method, similar to explicit calibration, requires gathering samples of estimated gaze points and their corresponding ground truth values. We observed that the user's interaction behavior provides calibration samples, as the location of the cursor at the moment the thumb is released can be regarded as the ground truth for the gaze point. Therefore, the user-unaware auto-calibration strategy continuously collects calibration data during interactions and dynamically adjusts the gaze estimation based on historical interaction samples.

Specifically, each time the user releases their thumb after completing an interaction, a calibration sample is recorded. Each calibration sample represents the position of the cursor at the moment the thumb is released, that is, the ground truth of the gaze point \( G_{gt}^i \), and the gaze point estimated by the neural network \( G_{E}^i \). The user-unaware auto-calibration aims to correct the current estimated gaze point \( G_{E}^t \) based on these samples, reducing the distance the user needs to drag and improving the interaction experience.

To achieve this, we initially designed an efficient auto-calibration strategy (Strategy 1) that fulfills our goals. Subsequently, we considered additional factors specific to mobile scenarios to further enhance the performance of the auto-calibration, leading to Strategy 2. The user study is discussed in detail in Section \ref{sec:study1}.

\subsubsection{Calibration Strategy 1}

In the \( t \)-th interaction, our strategy corrects the current estimated gaze point \( G_{E}^t \) by applying distance-inverse weighting to the offsets of historical samples \(\{{dG}^i | i = 1, 2, \dots, {t-1} \}\), resulting in the calibrated gaze point \( G_{C}^t \):
\begin{equation}
G_{C}^t = G_{E}^t + \sum_{i=1}^{t-1}{\lambda_{t}^{i} {dG}^i},
\end{equation}
where \( {dG}^i \) is the offset vector of the \(i\)-th sample, and \( \lambda_{t}^{i} \) is the distance-inverse weight of the \(i\)-th sample relative to the current prediction. These terms are defined as follows:
\begin{equation}
\left\{
\begin{aligned}
& {dG}^i = G_{gt}^i - G_{E}^i, \\
& \lambda_{t}^{i} = \frac{1}{S_{t} \lVert G_{gt}^i - G_{E}^i \rVert}.
\end{aligned}
\right.
\end{equation}

Here, \( \lVert G_{gt}^i - G_{E}^i \rVert \) represents the Euclidean distance between two points, and \( S_{t} = \sum_{j=1}^{t-1} \frac{1}{\lVert G_{gt}^j - G_{E}^j \rVert} \). This auto-calibration strategy effectively corrects the biases in the gaze estimation output by the network, which are influenced by user-specific and environment-specific factors.

\subsubsection{Calibration Strategy 2}

Head movements often affect the accuracy of gaze calibration. Therefore, we account for this factor in our auto-calibration strategy. Specifically, the user's estimated head pose relative to the camera is represented by a normalized direction vector \( H_E^i \). This information is obtained during the preprocessing stage of gaze estimation and retained in the calibration samples.

The calibrated gaze point \( G_{C}^t \) is determined with an additional weight \( h_{t}^{i} \) based on the head pose, and is expressed as:
\begin{equation}
G_{C}^t = G_{E}^t + \sum_{i=1}^{t-1}{\lambda_{t}^{i} h_{t}^{i} {dG}^i},
\end{equation}
where \( h_{t}^{i} = H_E^t \cdot H_E^i \) represents the cosine of the angle between the current head pose direction and the head pose direction of the \( i \)-th sample.

This calibration strategy is more robust to head movements relative to the phone, as demonstrated in our subsequent studies. Ultimately, our interaction technique adopts this strategy for user-unaware auto-calibration, reducing the average drag distance in GazeSwipe, thereby improving the user experience.

Our calibration strategies post-process the neural network output, weighting the results based on historical gaze points and head pose. This process does not involve updating the neural network's parameters. We found that biases caused by factors such as head pose changes are likely low-dimensional. Considering the real-time constraints of resource-limited mobile devices, post-processing alone delivers satisfactory performance.

\subsection{Gaze and Swipe Interaction}
\label{sec:interaction}

GazeSwipe leverages gaze for pointing at target elements, allowing the cursor to intuitively and quickly approach the target. It's important to note that previous studies have found that even with professional external eye trackers and additional calibration steps, gaze precision is not always sufficient for precise target selection. Therefore, manual adjustment of the cursor to fine-tune its position is necessary.

Our approach combines gaze with finger-swipe gestures, enabling users to control the gaze cursor through a seamless "touch-drag-release" motion. GazeSwipe displays the gaze cursor, which snaps to the nearest element in real-time on the screen. Users can tap anywhere on the screen with their thumb to lock the cursor, drag their thumb to adjust the cursor toward the target, and release it at the intended location to confirm the interaction. Additionally, if the gaze cursor is already accurately snapped to the target without requiring a swipe, a simple tap can confirm the interaction.

Efforts have been made to ensure GazeSwipe is smooth and user-friendly. Since gaze estimation errors increase with swipe distance, larger errors can make the operation more difficult and time-consuming. Therefore, incorporating gaze calibration methods to reduce estimation errors is essential. Our auto-calibration approach not only addresses this issue but also eliminates the need for explicit calibration processes. Moreover, the gaze cursor analyzes the layout and automatically snaps to the nearest element, bringing it closer to the target. These strategies make interactions more efficient and effortless.

GazeSwipe is activated and deactivated using a custom gesture. In our implementation, users can activate GazeSwipe by double-tapping the screen edge, which initiates gaze estimation and displays the gaze cursor. The same gesture can be used to deactivate it and return to the standard touchscreen mode.

To enhance the interaction experience and facilitate continuous operations, GazeSwipe remains compatible with standard touchscreen gestures while activated. Specifically, users can double-tap to trigger a short tap at the finger position or use a long swipe to scroll the page. This design emphasizes GazeSwipe’s ability to facilitate smooth interactions across different regions of the screen.

\section{Evaluation Design}

\subsection{Research Objectives}

In the forthcoming user study, participants will simulate typical usage scenarios with smartphones and tablets, enabling an evaluation of our method in an unconstrained environment. This evaluation will involve both quantitative and subjective comparisons. Our objective is to address the following research questions:

\textit{RQ1: How does the performance of mobile gaze estimation using front-facing cameras fare in interaction?} In the context of using built-in front-facing RGB cameras on handheld smartphones and tablets, we need to investigate the effectiveness of gaze estimation in our interaction method, which will be explored in Study 1.

\textit{RQ2: How do different calibration techniques affect the interaction experience?} Leveraging the insight that the interaction process itself serves as the calibration, we propose a user-unaware auto-calibration method. The effectiveness of this method compared to explicit calibration will be investigated in Study 1.

\textit{RQ3: How does the performance of GazeSwipe compare to other reachability solutions?} Compared to Direct Touch, One-Handed Mode, and Cursor-based Techniques without gaze input, can Gaze-Swipe provide a more seamless and accurate interaction experience in addressing reachability issues? Moreover, how does its performance differ between smartphones and tablets? These questions will be investigated in Study 2.

\subsection{Participants}
We recruited 15 participants (2 female, 13 male) aged between 21 and 29 years (\(M=24.33, SD=2.26\)). All of them was right-handed, and 12 of them wore glasses. Participants rated their experience with smartphone usage (\(M=4.87, SD=0.35\)), tablet usage (\(M=4.0, SD=1.0\)), and gaze interaction (\(M=3.53, SD=0.83\)) on a scale from 1 (no experience) to 5 (expert).

\subsection{Apparatus}
The user studies were conducted on two devices. For the smartphone, we used the Redmi K60, an Android device with a 6.67" screen, 1440×3200 px resolution, and a Qualcomm Snapdragon 8+ Gen1 SoC. For the tablet, we used the Xiaomi Pad 6, featuring an 11" screen, 2880×1800 px resolution, and a Snapdragon 870 SoC. The front cameras on both devices are positioned at the top center of the screen. Additionally, we tested the effectiveness of GazeSwipe on other Android devices with varying screen sizes, camera positions, and computational capabilities. Our method performed smoothly across all of these devices.

\begin{figure}[tb]
 \centering
 \includegraphics[width=\columnwidth]{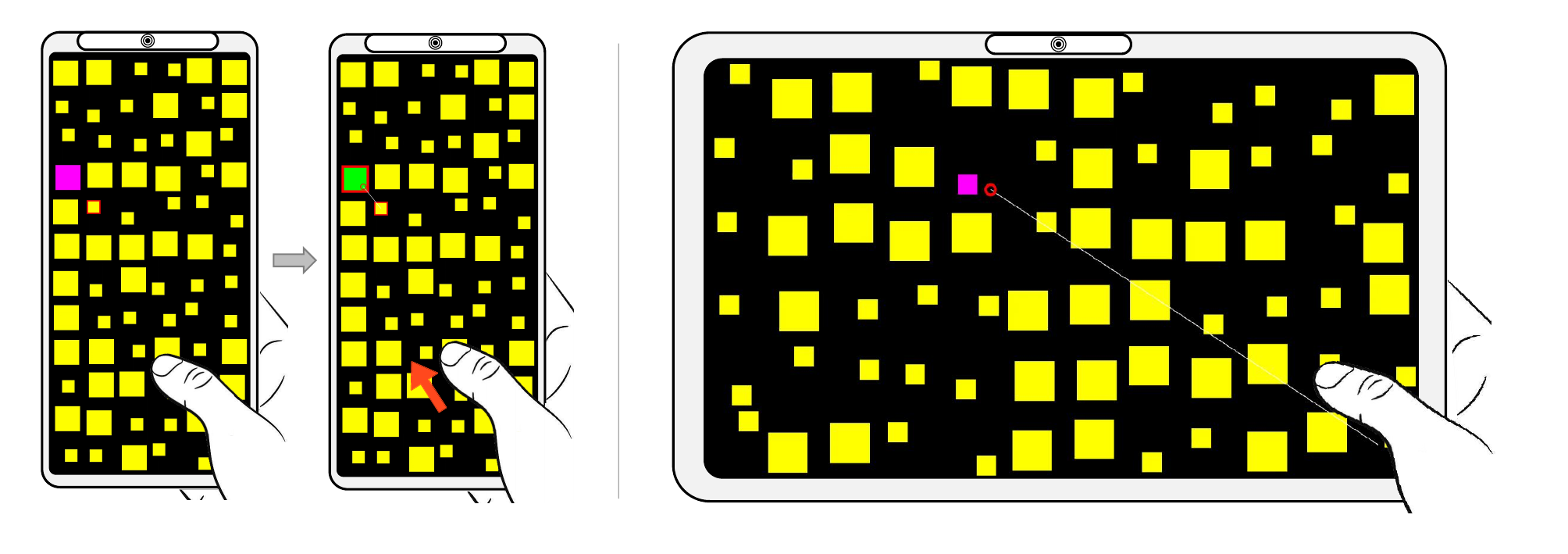}
 \caption{User interface during the studies. A 12×6 grid of square targets is randomly generated, comprising of two different sizes. Left: Smartphone interface (shown with GazeSwipe). Participants fixate their gaze on a target, causing the cursor to appear near it. They then swipe the cursor to the target, which turns green once reached. Right: Tablet interface (shown with Pure Cursor).}
 \Description{The user interface during the study shows a 12 by 6 grid of randomly generated square targets in two different sizes. On the left, the smartphone interface is displayed, demonstrating the GazeSwipe interaction technique. Participants fixate their gaze on a target, causing a cursor to appear nearby, and then use a swipe gesture to move the cursor toward the target. Once the target is reached, it changes to green. On the right, the tablet interface is displayed, showing a similar layout but designed for tablet use.}
 \label{fig:ui}
\end{figure}

\subsection{User Interface}
The user interface is illustrated in Fig. \ref{fig:ui}. In the user study, the smartphone screen, with a resolution of 2268×1080 pt, is divided into a 12×6 grid, where each cell measures 189×180 pt. Each grid cell contains a randomly positioned yellow square element, either 50×50 pt or 100×100 pt in size. During the experiment, one randomly selected element is highlighted in magenta, and participants are instructed to perform a short tap interaction on this target element using the specified interaction method.

For the Pure Cursor interaction method, a red circle represents the cursor. When the user's thumb touches the screen, the cursor initially appears at the touch point. As the thumb is swiped, both the starting cursor and the current cursor are displayed, with a directional line connecting them. Releasing the thumb triggers the interaction at the position of the current cursor. For GazeSwipe, the starting cursor automatically snaps to the nearest element, which is highlighted with a red outline around its edges. If the current cursor is positioned over the magenta target element, the element turns green.

\subsection{Evaluation Metrics}

\subsubsection{Objective Metrics}
In Study 1, we evaluated gaze estimation error across different calibration strategies, whereas in Study 2, we examined the latter three objective metrics across the four reachability techniques:

\textbf{Gaze Estimation Error.} This metric is determined by calculating the difference between the cursor coordinate when the user releases their thumb and the estimated gaze point. Gaze estimation error directly reflects the accuracy of gaze guidance, which in turn affects the drag distance during user interaction.

\textbf{Thumb Movement Distance.} This metric quantifies the distance traveled by the user's thumb from initial screen touch to release during each interaction.

\textbf{Completion Time.} This metric measures the duration from the user's initial thumb touch to its release during each interaction.

\textbf{Success Rate.} This metric quantifies the proportion of interactions where the user accurately taps within the designated target area, relative to the total number of interactions conducted.

\subsubsection{Subjective Metrics}
In Study 2, we also collected participant feedback through a questionnaire regarding their experiences with the four reachability techniques. The questionnaire assessed aspects such as ease of use, unreachability, grip stability, and fatigue. Participants also ranked their preferences for the four techniques.

\section{Study 1: Calibration Strategies}
\label{sec:study1}
In this study, our objective was to assess the efficacy of using the front-facing camera for gaze estimation in reachability interactions on smartphones and tablets, as well as to evaluate the accuracy of various gaze calibration strategies. To achieve this, participants were instructed to hold the mobile devices with one hand and perform experiments on the specified user interface, testing four calibration techniques on both the smartphone and tablet.

\subsection{Comparative Setups}
Participants interacted with four different calibration techniques for gaze-and-swipe interaction:

\textbf{No calibration (NC).} This configuration uses the output of the gaze estimation method without calibration, serving as the baseline to evaluate the fundamental capability of the neural network in targeting elements on smartphones and tablets.

\textbf{Explicit calibration (EC).} This involves an explicit 9-point calibration process before usage to imporve gaze estimation accuracy.

\textbf{Auto-calibration Strategy 1 (AC1).} This implements the user-unaware auto-calibration method proposed in Section \ref{sec:calibration}, collecting samples during interactions to automatically refine gaze estimation.

\textbf{Auto-calibration Strategy 2 (AC2).} This strategy integrates head pose factors into the auto-calibration process, providing more robust and accurate gaze calibration.

\subsection{Procedure}

The 15 participants followed this procedure for the study:

\textbf{Preparation Phase:} Participants first completed a personal information questionnaire, which collected details such as age, gender, handedness, use of glasses, and prior experience with smartphone operation, tablet operation, and eye-tracking interaction. Afterward, they were given a 2-minute period to freely familiarize themselves with the gaze-and-swipe interaction method.

\textbf{Experimental Phase:} Each participant was assigned to complete 4 calibration setups × 64 targets × 2 device types = 512 interaction trials. At the beginning of each experiment setup, the program cleared previous calibration data and randomly generated target layouts. Target selection during the experiment was randomized by the program. Participants were instructed to interact naturally and relaxedly with the targets according to the provided instructions. The duration of each participant's study session lasted approximately 30 minutes. Additionally, one participant was selected to conduct an extended experiment with 128 target selections on the smartphone, which was used to observe how gaze estimation error changes across more interactions for the four calibration strategies.

\begin{figure}[tb]
 \centering
 \includegraphics[width=\columnwidth]{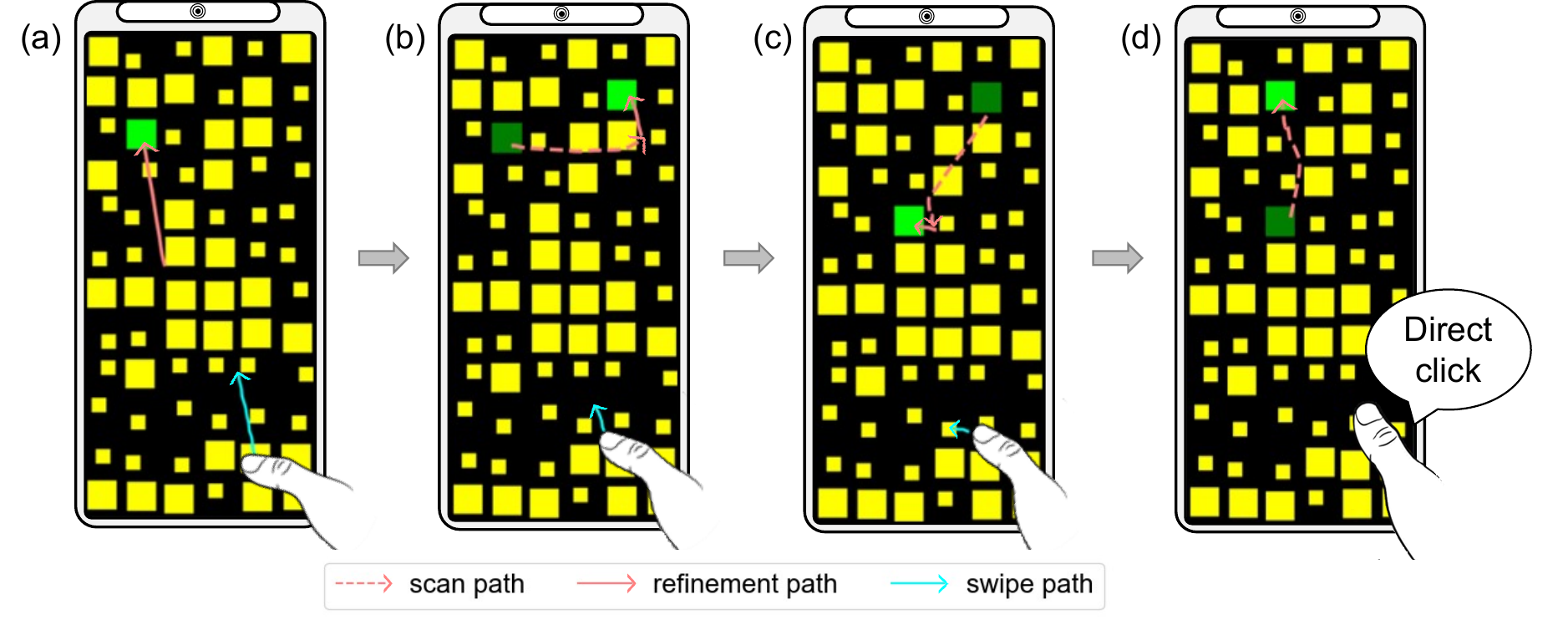}
 \caption{Example of gaze scanpaths collected during interactions. (a) At system startup, gaze errors are relatively large, requiring longer finger swipes. (b-c) After a few interactions, the length of finger swipes progressively decreases. (d) With prolonged use, the gaze can directly snap to the target, and a simple tap is sufficient to confirm the interaction.}
 \Description{The figure illustrates how gaze scanpaths evolve during the experiment, with gaze accuracy improving as usage progresses. In panel (a), at system startup, gaze errors are large, requiring longer finger swipes to correct the gaze. In panel (b-c), after a few interactions, the length of finger swipes decreases as gaze accuracy improves. In panel (d), after prolonged use, the gaze directly snaps to the target, allowing the interaction to be confirmed with a simple tap. }
 \label{fig:scanpath}
\end{figure}

\begin{figure}[tb]
 \centering
 \includegraphics[width=\columnwidth]{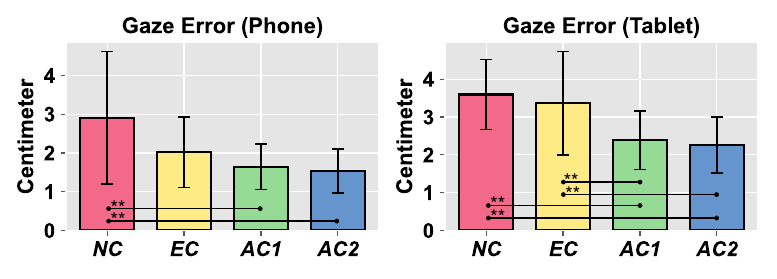}
 \caption{Gaze error for the four calibration strategies. Results on the smartphone (left) and on the tablet (right). The four strategies include: No Calibration (NC), Explicit Calibration (EC) using a 9-point calibration process, Auto-calibration Strategy 1 (AC1) without head pose, and Auto-calibration Strategy 2 (AC2) incorporating head pose. Error bars represent standard deviations, with significant differences marked by ** (\(p < 0.05\)).}
 \Description{The figure compares gaze estimation errors for four calibration strategies on smartphones and tablets. The left half of the figure displays results for smartphones, while the right half shows results for tablets. Error bars represent standard deviations for each calibration strategy. Significant differences between calibration strategies are marked with double asterisks (**), indicating a p-value less than 0.05. In both cases, the No Calibration (NC) strategy shows the highest error, followed by Eye Calibration (EC), Auto Calibration 1 (AC1), and Auto Calibration 2 (AC2), which has the lowest error.}
 \label{fig:study1-results}
\end{figure}

\subsection{Results}

The gaze estimation error for the four calibration strategies is illustrated in Fig. \ref{fig:study1-results}, with smartphone results on the left and tablet results on the right. For smartphones, a repeated-measures ANOVA revealed a significant effect of calibration strategy on gaze estimation error (\(F_{3,56} = 5.26, p < 0.005, \eta_p^2 = 0.22\)). Post hoc pairwise comparisons using the Tukey HSD method indicated that the mean gaze error followed the order NC>EC>AC1>AC2, with significant differences between NC-AC1 and NC-AC2 (\(p < 0.01\)). 

For tablets, a repeated-measures ANOVA test also showed a significant effect of calibration strategy on gaze estimation error (\(F_{3,56} = 7.06, p < 0.001, \eta_p^2 = 0.274\)). Post hoc pairwise comparisons using the Tukey HSD method revealed that the mean gaze error followed the same order (NC>EC>AC1>AC2), with significant differences observed between NC-AC1, NC-AC2, EC-AC1, and EC-AC2 (\(p < 0.05\)).

When comparing smartphones and tablets, the gaze error was consistently about 0.5 cm higher on tablets, both before and after calibration. This discrepancy may be attributed to the fact that tablet screens are typically viewed from a greater distance, resulting in lower resolution facial images for gaze estimation. Additionally, at the same angular error, a greater viewing distance on tablets translates into a larger projected error on the screen. Nevertheless, due to the larger screen and UI elements on tablets, several participants reported in the subsequent study that gaze pointing felt more accurate on tablets than on smartphones.

\begin{figure}[tb]
 \centering
 \includegraphics[width=\columnwidth]{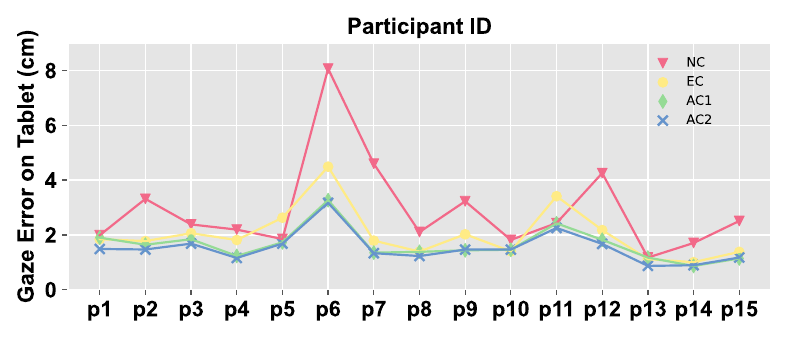}
 \caption{Gaze error for each participant. The error varies across participants before calibration but stabilizes after calibration, particularly with the two auto-calibration strategies.}
 \Description{The figure shows the gaze error for each participant under different calibration strategies. The gaze error varies significantly across participants before calibration, with some participants, such as P6, showing high error rates under the No Calibration strategy. After applying the auto-calibration strategies, the gaze error is reduced for most participants, with a noticeable decrease in error for P6, where the error drops to less than half of the original value. This highlights the effectiveness of the auto-calibration strategies in stabilizing gaze accuracy.}
 \label{fig:study1-participant}
\end{figure}

We also observed differences in gaze error among individuals under the four different calibration strategies. As shown in Fig. \ref{fig:study1-participant}, Participant P6 exhibited significant gaze error under No Calibration, but after applying the auto-calibration strategies, the gaze error reduced to less than half of the original value. This demonstrates the effectiveness and robustness of our calibration strategy.

\begin{figure}[tb]
 \centering
 \includegraphics[width=\columnwidth]{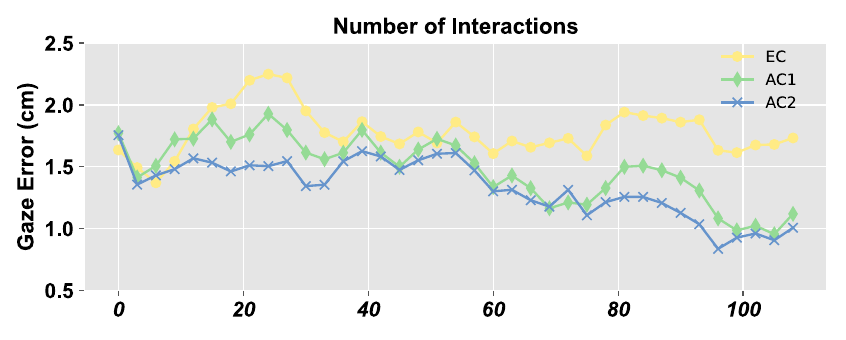}
 \caption{Gaze error of one participant over the number of interactions, with a sliding window of 16 and a step size of 4. As the number of interactions increases, AC1 and AC2 consistently collect the latest calibration samples, resulting in lower gaze errors compared to EC. Additionally, AC2, which accounts for head pose, demonstrates even lower gaze errors.}
 \Description{The figure displays the gaze error of one participant plotted against the number of interactions, using a sliding window of 16 and a step size of 4. The gaze error decreases as the number of interactions increases, especially for the two auto-calibration strategies (AC1 and AC2). AC1 and AC2 consistently collect recent calibration samples, reducing gaze error more effectively compared to explicit calibration (EC). AC2, which accounts for head pose, shows the lowest gaze errors over time. The no calibration (NC) strategy maintains a consistently high error of approximately 4 cm, which is not included in the graph.}
 \label{fig:study1-errorovertime}
\end{figure}

As illustrated in Fig. \ref{fig:study1-errorovertime}, we examined the variation of gaze error over the number of interactions for the four calibration strategies. NC consistently maintained high gaze error (around 4 cm, not plotted in the figure). Although EC initially had lower gaze error compared to the auto-calibration methods, both AC1 and AC2 eventually outperformed EC as the number of interactions increased. The gaze error of EC fluctuated within a certain range, which we attribute to changes in user posture over time. These posture changes could lead to deviations from the initial calibration conditions, affecting the accuracy of gaze estimation. In contrast, AC1 and AC2 automatically recalibrated using the latest interaction samples, stabilizing gaze error at relatively low levels after a certain number of interactions. Since AC2 benefits from posture information in historical calibration samples, it was able to interpolate calibration offsets based on similar postures in the past. This enabled AC2 to generally achieve lower gaze error than AC1.

\subsection{Discussion}

In Study 1, our results addressed the first two research questions. Regarding RQ1, we observed substantial gaze errors in the uncalibrated output of the gaze estimation method for certain users, with some individual error distances approaching the width of the screen, indicating that the uncalibrated method could not consistently provide reliable gaze pointing. For RQ2, we found that applying calibration strategies significantly improved the quality of gaze estimation. Moreover, the auto-calibration strategies outperformed the explicit 9-point calibration method, as they continuously accumulated new samples during interaction to enhance the most recent results. Additionally, between the two auto-calibration strategies, the results indicated an advantage of AC2 over AC1. Therefore, we selected AC2 as the calibration strategy for GazeSwipe.

\section{Study 2: Reachability Techniques}

In this study, we quantitatively evaluated the effectiveness of Gaze-Swipe in addressing reachability issues on mobile touchscreens compared to other techniques. We selected Strategy 2 as the auto-calibration method for GazeSwipe and compared it with three other techniques: Direct Touch, One-handed Mode, and Pure Cursor.

\subsection{Comparative Setups}
We compared the following four techniques on the smartphone, and since the thumb cannot directly touch certain areas when holding the tablet with one hand, we compared only the latter three techniques on the tablet:

\textbf{Direct Touch (DT).} While holding the phone with one hand, users’ thumbs typically reach only a portion of the screen, but they can adjust their grip to touch otherwise unreachable areas. We include this Direct Touch method as a baseline technique.

\textbf{One-handed Mode (OM).} This mode shrinks the user interface to a size fully reachable by the thumb and displays it in the bottom right corner of the screen (or bottom left for left-handed users). This feature allows users to directly touch areas that would normally be outside their thumb's reach and is commonly available on many Android smartphones.

\textbf{Pure Cursor (PC).} A cursor-based technique without gaze guidance, similar to BezelCursor \cite{li_bezelcursor_2013}, is employed. The main difference between this method and GazeSwipe is the initial position of the cursor. Here, the cursor starts at the user's touch position and extends linearly along the direction of the swipe by a certain factor, triggering interaction upon releasing the thumb. 

\textbf{GazeSwipe (GS).} Our proposed technique utilizes user-unaware auto-calibration Strategy 2 to deliver robust gaze
estimation results.

In the study, we randomized the order in which participants experienced each technique to ensure fairness in the comparison.

\begin{figure*}[tb]
 \centering
 \includegraphics[width=\linewidth]{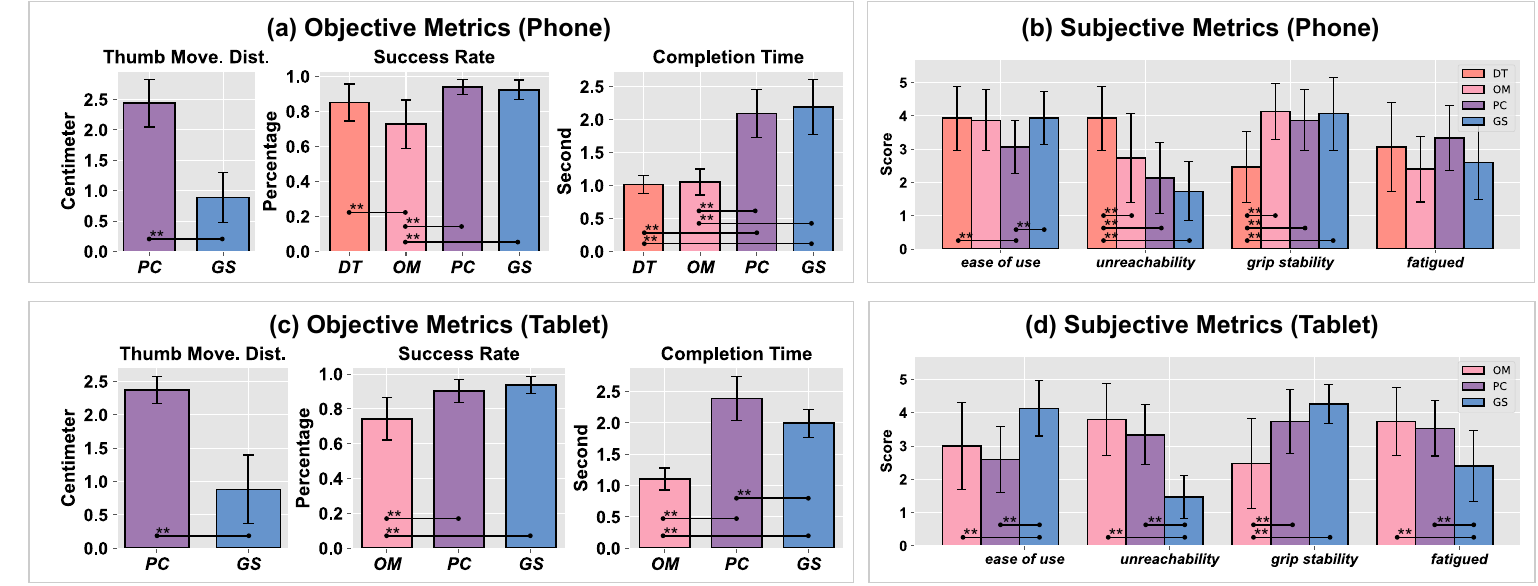}
 \caption{Objective and subjective results of Study 2. Panels (a-b) depict the results on the smartphone, while panels (c-d) illustrate the results on the tablet. The four techniques include: Direct Touch (DT), One-handed Mode (OM), Pure Cursor (PC), and GazeSwipe (GS). Error bars denote the standard deviation, with significant differences indicated by ** (\(p < 0.05\)).}
 \Description{This figure presents both objective and subjective results of Study 2 for smartphone and tablet interactions. Panels (a) and (b) show the results on the smartphone, while panels (c) and (d) display the results on the tablet. Error bars indicate standard deviation. Significant differences between techniques are marked with double asterisks (p < 0.05). Various metrics, including thumb movement distance, success rate, ease of use, unreachability, grip stability, and fatigue, are compared across different interaction techniques such as Direct Tapping (DT), Offset Mapping (OM), Predictive Cursor (PC), and Gaze Swipe (GS). Detailed analysis is provided in the text.}
 \label{fig:study2-results}
\end{figure*}

\begin{figure}[tb]
 \centering
 \includegraphics[width=\columnwidth]{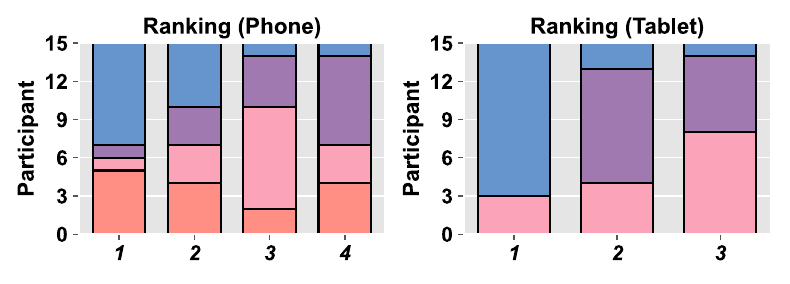}
 \caption{Preference ranking of the four reachability techniques on both smartphone and tablet (tablet excludes DT). A rank of 1 indicates the most preferred method. It can be seen that GazeSwipe (GS) received the highest preference on both devices.}
 \Description{This figure shows the preference ranking of four reachability techniques, GazeSwipe (GS), Predictive Cursor (PC), Offset Mapping (OM), and Direct Tapping (DT), on both smartphone and tablet. On the tablet, DT is excluded. A rank of 1 represents the most preferred method. GS is ranked as the highest preferred method across both devices.}
 \label{fig:study2-ranking}
\end{figure}

\subsection{Procedure}
\label{sec:study2-procedure}
Each participant was required to go through the following process:

\textbf{Preparation Phase:} After completing the personal information collection, participants were given 2 minutes to freely practice the four interaction techniques.

\textbf{Experimental Phase:} Each participant completed 4 techniques × 32 targets × 2 device types = 256 trials. Similar to the experimental phase in Study 1, the program randomly generated target layouts and selected targets for each trial. Participants were instructed to interact naturally with the targets using the specified techniques, simulating their typical smartphone usage. \mytextcolor{blue}{The calibration data from the preparation phase was cleared before the start of this phase, ensuring that no pre-calibration data was used in GazeSwipe.} Participants were allowed to take short breaks after completing a set of trials if desired, and the total experiment duration for each participant was within 30 minutes. 

\textbf{Subjective Evaluation:} After completing a set of trials for each technique, participants were asked to rate their level of agreement on a Likert scale (1: totally disagree; 5: totally agree) with statements including ``The technique was easy to apply'', ``I had difficulty tapping on certain areas of the screen'', ``I maintained a stable grip during the interaction'', and ``I felt fatigued after using this technique for a while''. Upon completing all experiments, they were asked to rank their preferences for the four techniques. Since our study specifically focuses on mobile touchscreen reachability, we prioritize metrics that reflect system usability and user workload based on the task context. Similar subjective questionnaires and metrics have been employed in several prior studies \cite{corsten_forceray_2019, voelker_headreach_2020}.

\subsection{Results on the Smartphone}

Our interest is to understand the impact of various reachability solution techniques on user performance. We first analyze user performance on the smartphone. We conducted repeated-measures ANOVA and post hoc pairwise comparisons to evaluate and analyze the objective and subjective metrics recorded in Study 2, including thumb movement distance, success rate, completion time, ease of use, unreachability, grip stability, fatigue, and preference ranking.

\subsubsection{Objective Measures}

The thumb movement distance is shown in the left column of Fig. \ref{fig:study2-results} (a). Since DT and OM do not involve swiping, this metric is not applicable to them. Thus, we only compare PC and GS. Repeated-measures ANOVA revealed a significant effect of technique on \textbf{thumb movement distance} (\(F_{1,28} = 113.567, p < 0.001, \eta_p^2 = 0.802\)). Tukey HSD post hoc pairwise comparisons indicated significant differences between the two techniques (\(p < 0.001\)). We found that PC requires a significantly longer thumb movement distance (2.436 cm) compared to GS (0.887 cm). This indicates that gaze pointing greatly reduces the swiping distance required for users, alleviating user fatigue and enabling more precise target selection.

The success rate is shown in the center column of Fig. \ref{fig:study2-results} (a). Technique has a significant impact on \textbf{success rate} (\(F_{3,56} = 16.094, p < 0.001, \eta_p^2 = 0.463\)). Tukey HSD post hoc pairwise comparisons indicated significant differences between DT-OM, OM-PC, and OM-GS (\(p < 0.001\)). The results indicate that, compared to DT (\(85.1\%\)) and OM (\(72.6\%\)), the success rates of PC (\(93.9\%\)) and GS (\(92.3\%\)) are significantly higher. 

The completion time is shown in the right column of Fig. \ref{fig:study2-results} (a). Technique has a significant impact on \textbf{completion time} (\(F_{3,56} = 67.832, p < 0.001, \eta_p^2 = 0.784\)). Tukey HSD post hoc pairwise comparisons are all significant (\(p < 0.001\)) except DT-OM and PC-GS. As expected, DT (1.016 s) and OM (1.054 s) are the fastest due to their lack of swiping process, and participants' familiarity with the direct tapping interaction. However, as mentioned earlier, their success rate is significantly lower. PC (2.091 s) ranks third in terms of completion time, while GS (2.190 s) takes the longest, but the difference in completion time between GS and PC is insignificant.

\begin{figure}[tb]
 \centering
 \includegraphics[width=\columnwidth]{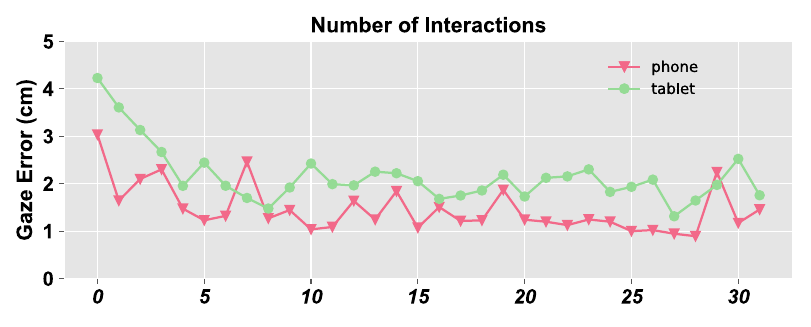}
 \caption{\mytextcolor{blue}{Mean gaze error of all participants over the number of interactions (no sliding window) in Study 2 for both phone and tablet. The gaze error is relatively high at the beginning, but decreases to a lower level as the number of interactions increases on both devices.}}
 \Description{The figure displays the mean gaze error of all participants plotted against the number of interactions in Study 2, with no sliding window, for both phone and tablet. Initially, the gaze error is relatively high, approximately 3 cm for the phone and 4 cm for the tablet, but it decreases as the number of interactions increases on both devices, reaching a lower and more stable level over time, around 1.5 cm for the phone and 2 cm for the tablet. The gaze error decreases more significantly as participants continue interacting with the devices, indicating improved accuracy with repeated use.}
 \label{fig:study2-errorovertime}
\end{figure}

\subsubsection{Subjective Measures}

Fig. \ref{fig:study2-results} (b) illustrates the results of the user survey. The four indicators correspond to the four survey questions presented in Section \ref{sec:study2-procedure}. Repeated-measures ANOVA shows that technique has a significant effect on \textbf{ease of use} (\(F_{3,56} = 3.54, p < 0.05, \eta_p^2 = 0.159\)), and post hoc pairwise comparisons reveal significant differences between DT-PC and PC-GS (\(p < 0.05\)), with the sequence DT>GS>OM>PC. Technique also has a significant effect on \textbf{unreachability} (\(F_{3,56} = 11.975, p < 0.001, \eta_p^2 = 0.391\)). Significant differences were found between DT-OM, DT-PC, and PC-GS, with the order GS<PC<OM<DT, suggesting that DT suffers from severe reachability issues. In addition, technique has a significant effect on \textbf{grip stability} (\(F_{3,56} = 9.586, p < 0.001, \eta_p^2 = 0.339\)). Significant differences were found between DT-OM, DT-PC, and DT-GS, indicating that users felt significantly less stable when using DT, while the other three methods provided better grip stability. However, technique did not have a significant effect on \textbf{fatigue} (\(F_{3,56} = 2.199, p>0.05, \eta_p^2 = 0.105\)). For the \textbf{preference ranking}, users' average preferences are ranked as GS>DT>OM>PC, as shown in Fig. \ref{fig:study2-ranking}.

\subsection{Results on the Tablet}

We then evaluated the metrics on the tablet. Due to the significant difference in screen size between the tablet and the smartphone, the metrics exhibited distinct characteristics.

\subsubsection{Objective Measures}

Fig. \ref{fig:study2-results} (c) illustrates the results of objective measures on the tablet. Repeated-measures ANOVA revealed a significant effect of technique on \textbf{thumb movement distance} (\(F_{1,28} = 109.211, p < 0.001, \eta_p^2 = 0.796\)). Tukey HSD post hoc pairwise comparisons indicated significant differences between the two techniques (\(p < 0.001\)). Similar to the smartphone results, PC requires a significantly longer thumb movement distance (2.372 cm) compared to GS (0.882 cm). Technique also has a significant impact on \textbf{success rate} (\(F_{2,42} = 22.359, p < 0.001, \eta_p^2 = 0.516\)). Tukey HSD post hoc pairwise comparisons revealed significant differences between OM-PC and OM-GS (\(p < 0.001\)). The results indicate that, compared to OM (\(74.3\%\)), the success rates of PC (\(90.1\%\)) and GS (\(93.8\%\)) are significantly higher. Due to the larger screen size of the tablet, PC requires a longer cursor displacement with the same thumb movement distance, which increases the likelihood of errors. Technique has a significant impact on \textbf{completion time} (\(F_{2,42} = 96.123, p < 0.001, \eta_p^2 = 0.821\)). Tukey HSD post hoc pairwise comparisons are all significant (\(p < 0.001\)). OM (1.107 s) remains the fastest, while GS (1.993 s) and PC (2.391 s) are slower.

\subsubsection{Subjective Measures}

Fig. \ref{fig:study2-results} (d) illustrates the results of subjective measures on the tablet. Repeated-measures ANOVA shows that technique has a significant effect on \textbf{ease of use} (\(F_{2,42} = 8.42, p < 0.001, \eta_p^2 = 0.286\)). Post hoc pairwise comparisons reveal significant differences between OM-GS and PC-GS (\(p < 0.05\)), with the ranking of GS>OM>PC. Technique also significantly impacts \textbf{unreachability} (\(F_{2,42} = 28.697, p < 0.001, \eta_p^2 = 0.577\)). Significant differences were found between OM-GS and PC-GS, with the order GS<PC<OM, indicating that GS has better reachability compared to the other two methods. Additionally, technique significantly affects \textbf{grip stability} (\(F_{2,42} = 12.352, p < 0.001, \eta_p^2 = 0.37\)). Significant differences were found between OM-PC and OM-GS, with the ranking GS>PC>OM, suggesting that PC and GS offer better grip stability than OM. Technique also significantly influences \textbf{fatigue} (\(F_{2,42} = 8.089, p < 0.001, \eta_p^2 = 0.278\)), with the order GS<PC<OM. Significant differences were observed between OM-GS and PC-GS, indicating that GS is notably less fatiguing on the tablet. For \textbf{preference ranking}, users' average preferences are ranked as GS>OM>PC, as shown in Fig. \ref{fig:study2-ranking}.

\subsection{Discussion}

Following the completion of Study 2, we discussed RQ3, comparing the performance differences between our proposed GazeSwipe technique (GS) and the other three techniques: DT, OM, and PC.

\subsubsection{Direct Touch vs. GazeSwipe}
Direct Touch (DT) serves as the baseline in this study, exhibiting significant reachability issues as evidenced by participants' subjective feedback, indicated by higher unreachability ratings. This necessitates frequent grip adjustments, leading to feelings of instability and fatigue, along with a lower success rate than cursor-based method. However, DT still ranked second in preference on smartphones, just after GazeSwipe. This may be attributed to its familiar operation and faster completion times. 

\subsubsection{Pure Cursor vs. GazeSwipe}
Pure Cursor (PC) addresses DT's reachability issues using cursor manipulation, which proves to be an effective solution. Results show higher success rates and a perceived sense of greater reachability and grip stability compared to DT. However, PC's cursor extends from the thumb touch point, requiring longer thumb movement distances and resulting in higher fatigue compared to GazeSwipe. On tablets, controlling the cursor across the large screen with small thumb movements proves even more challenging. As a result, PC received the lowest preference ranking on both devices.

\subsubsection{One-handed Mode vs. GazeSwipe}
One-handed Mode (OM) tackles reachability issues by shrinking the screen, a feature employed in many Android smartphones. On smartphones, the results show that OM and GazeSwipe receive similar ratings in ease of use, grip stability, and fatigue. However, OM falls short of both GazeSwipe and even DT in terms of success rate, as the reduced element size makes thumb tapping more difficult. OM does have an advantage in completion time over GazeSwipe, as it, like DT, does not require swiping.

On tablets, GazeSwipe demonstrates a more pronounced advantage over OM, outperforming it in all metrics except for completion time. This is likely due to the large screen requiring a greater reduction in interface size for OM, making precise taps harder. On the other hand, enlarging the interface to improve accuracy leads to reduced reachability and grip stability, increasing fatigue.

Interestingly, when we asked the 15 participants whether they prioritized completion time or success rate during interactions, 11 preferred success rate, while 4 considered both equally important. This suggests that users may tolerate moderately longer completion times if the interaction method offers greater precision.

\subsubsection{Smartphone vs. Tablet}
When comparing the performance of GazeSwipe on smartphones and tablets, some participants reported that the interaction felt more effective on tablets, with several attributing this to more accurate gaze pointing on the larger screen. Others noted that, in contrast, One-handed Mode and Pure Cursor were less ideal solutions on tablets.

From the objective metrics, although the gaze estimation error after calibration was slightly higher on tablets compared to smartphones, the larger UI elements on tablets allowed for more precise gaze-based target selection. Additionally, the success rate and completion time for GazeSwipe were similar on both devices, with a slight advantage observed on tablets.

Subjectively, GazeSwipe did not show significant differences compared to One-handed Mode across the four subjective metrics on smartphones. However, on tablets, GazeSwipe significantly outperformed One-handed Mode in all four metrics, indicating that it provides the best reachability experience on tablets.

In terms of preference ranking, GazeSwipe was the most favored solution on both smartphones and tablets, highlighting its intuitive and seamless interaction approach as a effective option across both device types.

\subsubsection{Overall}
In summary, GazeSwipe effectively addresses the reachability issues inherent in Direct Touch, significantly enhancing user experience compared to Pure Cursor. Furthermore, GazeSwipe outperforms or matches One-handed Mode in all aspects except for completion time, with this advantage being even more significant on tablets.

\section{Limitations, Future Work and Conclusion}
\subsection{Limitations and Future Work}

While our method demonstrates effectiveness, it still has some limitations. \mytextcolor{blue}{Firstly, since GazeSwipe does not rely on explicit calibration, the gaze point may be less accurate initially, although it improves in accuracy after several interactions. Secondly,} due to the use of visual methods, our technique may not be suitable for dark environments. This could potentially be improved by implementing automatic screen brightness adjustment during interaction to provide adequate illumination. \mytextcolor{blue}{Thirdly,} some participants observed that the starting position of the gaze-guided cursor could be located anywhere around the target, leading to varying drag directions with each interaction, which might increase their response time. This could possibly be addressed by introducing an offset to the initial position to maintain consistent drag directions. \mytextcolor{blue}{Furthermore,} our method lacks user recognition and does not store historical calibration information, which could limit its usability in practical applications. \mytextcolor{blue}{Lastly,} for more complex applications, such as typing, the interaction method remains too slow to be practical in real-world scenarios.

As the combination of gaze, finger swipe, and auto-calibration provides a robust solution for reachability issues on mobile devices, future work could focus on further improving the accuracy of gaze estimation or designing more efficient auto-calibration strategies. Potential directions for enhancing auto-calibration strategies include making better use of additional samples collected during a single interaction, seamlessly integrating the gaze estimation network with the calibrator, and considering additional factors such as time, lighting, and other environmental variables. Furthermore, exploring more multimodal interaction methods for daily use on mobile devices presents a promising avenue for future research.

\subsection{Conclusion}

In this paper, we propose GazeSwipe, a novel gaze-and-swipe interaction technique in both smartphones and tablets aimed at addressing reachability issues encountered when using mobile touchscreens with one hand. Users simply need to naturally gaze at the desired target, while controlling the gaze cursor to reach the target through a finger-swipe operation at any position on the screen. To ensure robust gaze estimation results, especially when head movements occur frequently, we introduce a user-unaware auto-calibration strategy. Based on the insight that the interaction process can serve as the calibration process, this method eliminates the need for additional calibration procedures, enabling dynamic and implicit calibration. We also investigated the effectiveness of our auto-calibration strategy and interaction technique. The results indicate that our auto-calibration provides more accurate gaze pointing compared to no calibration and 9-point calibration and that our method achieves high success rates and top preference rankings against Direct Touch, One-handed Mode and Pure Cursor on both smartphones and tablets. Specifically on tablets, our method outperforms others across all subjective and objective metrics, with the exception of a moderately longer completion time.

\begin{acks}
This work was supported by the Beijing Natural Science Foundation under grant No. L242019.
\end{acks}

\bibliographystyle{ACM-Reference-Format}
\bibliography{sample-base}

\end{document}